\documentclass[acmsmall,screen,nonacm]{acmart}
\makeatletter
\def\@ACM@journal@bibstrip{} 
\makeatother
\AtBeginDocument{%
  }

\usepackage{amsmath,amssymb,amsfonts}
\usepackage{graphicx}
\usepackage{forest}
\usepackage{subcaption}
\usepackage{textcomp}
\usepackage{tikz}
\usepackage{xcolor}
\usepackage{multirow}
\usepackage{algorithm}
\usepackage{algpseudocode}
\usepackage{pgfplots}
\usepackage{amsmath} 
\pgfplotsset{compat=1.18} 
\usepackage{lipsum} 
\usepackage{booktabs} 
\usepackage{siunitx}  
\usepackage{rotating} 
\usepackage{tabularx}
\usepackage{multirow}
\usepackage{geometry}
\usepackage{listings}

\usepackage{tcolorbox}
\tcbuselibrary{listings}

\definecolor{codegreen}{rgb}{0,0.6,0}
\definecolor{codegray}{rgb}{0.5,0.5,0.5}
\definecolor{codepurple}{rgb}{0.58,0,0.82}
\definecolor{backcolour}{rgb}{0.95,0.95,0.92}

\lstdefinestyle{mystyle}{
    backgroundcolor=\color{backcolour},
    commentstyle=\color{codegreen},
    keywordstyle=\color{magenta},
    numberstyle=\tiny\color{codegray},
    stringstyle=\color{codepurple},
    basicstyle=\ttfamily\footnotesize,
    breakatwhitespace=false,
    breaklines=true,
    captionpos=b,
    keepspaces=true,
    numbers=left,
    numbersep=5pt,
    showspaces=false,
    showstringspaces=false,
    showtabs=false,
    tabsize=2
}

\setcopyright{acmlicensed}
\copyrightyear{2026}
\acmYear{2026}
\acmDOI{XXXXXXX.XXXXXXX}
\acmConference[FSE'26]{ACM International Conference on the Foundations of Software Engineering}{xxx,
  2026}{xxx}
\acmISBN{978-1-4503-XXXX-X/2018/06}

\begin{document}

\title{Synergy-Guided Compiler Auto-Tuning of Nested LLVM Pass Pipelines}

\author{Haolin Pan}
\affiliation{%
  \institution{Hangzhou Institute for Advanced Study at UCAS}
  \city{Hangzhou}
  \country{China}
}
\affiliation{%
  \institution{Institute of Software, Chinese Academy of Sciences}
  \city{Beijing}
  \country{China}
}
\affiliation{%
  \institution{University of Chinese Academy of Sciences}
  \city{Beijing}
  \country{China}
}
\email{panhaolin21@mails.ucas.ac.cn}

\author{Jinyuan Dong}
\affiliation{%
  \institution{Institute of Software, Chinese Academy of Sciences}
  \city{Beijing}
  \country{China}
}
\affiliation{%
  \institution{University of Chinese Academy of Sciences}
  \city{Beijing}
  \country{China}}
\email{dongjinyuan24@mails.ucas.ac.cn}

\author{Mingjie Xing}
\authornote{Corresponding Author.}
\affiliation{%
  \institution{Institute of Software, Chinese Academy of Sciences}
  \city{Beijing}
  \country{China}}
\email{mingjie@iscas.ac.cn}

\author{Yanjun Wu}
\affiliation{%
  \institution{Institute of Software, Chinese Academy of Sciences}
  \city{Beijing}
  \country{China}}
\affiliation{%
  \institution{University of Chinese Academy of Sciences}
  \city{Beijing}
  \country{China}}
\email{yanjun@iscas.ac.cn}

\renewcommand{\shortauthors}{Haolin Pan et al.}

\begin{abstract}
Compiler optimization relies on sequences of passes to improve program performance. Selecting and ordering these passes automatically, known as compiler auto-tuning, is challenging due to the large and complex search space. Existing approaches generally assume a linear sequence of passes, a model compatible with legacy compilers but fundamentally misaligned with the hierarchical design of the LLVM New Pass Manager. This misalignment prevents them from guaranteeing the generation of syntactically valid optimization pipelines. In this work, we present a new auto-tuning framework built from the ground up for the New Pass Manager. We introduce a formal grammar to define the space of valid nested pipelines and a forest-based data structure for their native representation. Upon this foundation, we develop a structure-aware Genetic Algorithm whose operators manipulate these forests directly, ensuring that all candidate solutions are valid by construction. The framework first mines synergistic pass relationships to guide the search. An optional refinement stage further explores subtle performance variations arising from different valid structural arrangements.

We evaluate our approach on seven benchmark datasets using LLVM 18.1.6. The discovered pipelines achieve an average of 13.62\% additional instruction count reduction compared to the standard \texttt{opt -Oz} optimization level, showing that our framework is capable of navigating this complex, constrained search space to identify valid and effective pass pipelines.
\end{abstract}

\ccsdesc[500]{Software and its engineering~Compilers}
\ccsdesc[300]{Software and its engineering~Search-based software engineering}
\ccsdesc[300]{Computing methodologies~Machine learning}

\keywords{Compiler auto-tuning, Code optimization, Knowledge base, Pass sequence tuning}


\maketitle

\section{Introduction}
\label{chap:introduction}

Modern compilers use many optimization passes to improve program performance. Selecting and ordering these passes to form an effective sequence is a central task in compiler engineering. Standard optimization levels such as \texttt{-Oz} offer a general-purpose solution, but they are often not optimal for specific applications. Compiler auto-tuning aims to reduce this gap by automatically finding program-specific pass sequences. The main challenge is the very large search space of possible combinations. This difficulty has motivated research into heuristic search and machine learning approaches that can discover better customized strategies.

Most existing auto-tuning research focused on legacy compiler infrastructures, such as the LLVM legacy Pass Manager, where optimization pipelines are modeled as linear sequences. In this setting, the key challenge is the \textbf{combinatorial explosion}: for $N$ passes and a sequence of length $k$, the number of possible permutations grows factorially, which makes exhaustive search infeasible. To address this, many studies explored efficient search strategies in the sequential space. Early work in \textbf{iterative compilation (IC)}~\cite{bodin1998iterative, opentuner} showed the effectiveness of empirical search guided by heuristics such as Genetic Algorithms~\cite{GA}, though with high compilation cost. Later work introduced \textbf{machine learning-based methods}~\cite{leather2020machine}, which built predictive models linking program features to useful optimizations. These included supervised learning to predict pass benefits~\cite{Comptuner} and reinforcement learning agents that treat pass ordering as a sequential decision process~\cite{autophase, CompilerGym}. More recently, \textbf{Large Language Models (LLMs)} have been applied to learn optimization patterns directly from code~\cite{cummins2024metalargelanguagemodel}. Although diverse in methodology, these approaches share the simplifying assumption of a flat, sequential pipeline, which no longer reflects current compiler infrastructures.

The LLVM New Pass Manager (New PM) introduces a hierarchical design that reflects the structure of the Intermediate Representation (\textbf{Module > CGSCC > Function > Loop}). This architectural shift renders simple linear pass sequences not just insufficient, but \textbf{invalid}. Any auto-tuner for the New PM faces a primary, non-negotiable requirement: it \textbf{must} produce syntactically valid nested pipelines that conform to the New PM's grammar. Consequently, existing approaches that manipulate flat sequences are unusable in this new context, as they lack the fundamental mechanisms to ensure structural validity and navigate the constrained search space.

This fundamental requirement motivates our work: to create a complete auto-tuning framework built from the ground up for the LLVM New PM. To achieve this, we first establish a formal foundation for the problem. We introduce a grammar that precisely defines the space of all valid hierarchical pipelines and propose a logical forest data structure for their direct in-memory representation. This representation is the cornerstone of our structure-aware Genetic Algorithm. Unlike traditional approaches that manipulate linear sequences, its genetic operators (e.g., subtree crossover and mutation) operate directly on the forest structure itself. This design choice provides a guarantee: every candidate pipeline generated during the evolutionary search is, by construction, syntactically valid. To navigate this vast space of valid pipelines efficiently, the search is guided by a knowledge base of synergistic pass relationships mined offline. Furthermore, the framework leverages its deep structural awareness through an optional refinement stage, enabling it to explore small performance nuances between different valid nesting structures.

We implemented and evaluated our framework on seven benchmark datasets against the \texttt{opt -Oz} optimization level of LLVM 18.1.6. Our pipelines achieve an average of \textbf{13.62\%} additional reduction in instruction count compared to \texttt{opt -Oz}, and they perform better than existing auto-tuning methods.

This paper makes the following contributions:
\begin{itemize}
    \item We provide the first formalization of the auto-tuning problem for the LLVM New Pass Manager's hierarchical architecture. We introduce a grammar to describe the space of valid nested pipelines and propose a logical forest data structure for their representation and manipulation.
    
    \item We propose a complete methodology for discovering high-quality pass sequences that is aware of structural constraints. This includes a new \textbf{synergy mining technique} to build a knowledge graph of structured pass relationships, and a \textbf{Synergy-Guided Genetic Algorithm} that uses this graph to construct high-performance hierarchical pass sequences.

    \item We identify and analyze the performance impact of different valid nesting structures for a fixed pass sequence. We then introduce a targeted refinement mechanism that leverages this insight to capture modest but consistent performance improvements, demonstrating the ability to exploit structural variations for additional optimization.

    \item We evaluate our framework and show that it achieves \textbf{superior instruction count reduction compared to existing auto-tuning methods} on the LLVM Intermediate Representation. Our method consistently outperforms LLVM's standard optimization \texttt{opt -Oz} levels across a wide set of benchmarks.

\end{itemize}

\section{Background and Motivation}
\label{chap:background_motivation}

\subsection{Background}
\label{ssec:background}

The LLVM compiler infrastructure's New PM constitutes an architectural change from the earlier linear optimization pipeline. It organizes transformations and analyses in a hierarchical manner, supporting module-level, function-level, CGSCC-level and loop-level passes within a unified framework. The design provides mechanisms for finer-grained control of pass execution, improved analysis reuse, and support for parallelism in both analysis and transformation. These features reflect a response to the growing scale and complexity of modern software and hardware, and they establish a foundation for exploring compiler optimization and auto-tuning in a more flexible setting.

A key characteristic of the New PM is its replacement of the conventional flat-list pipeline with a hierarchical organization that reflects the nested structure of the IR. This hierarchy comprises four principal levels of IR units: \textbf{Module}, the top-level container for a translation unit; \textbf{CGSCC} (Call Graph Strongly Connected Component), a unit designed for inter-procedural optimizations; \textbf{Function}, a single function body; and \textbf{Loop}, a natural loop within a function's control-flow graph. Each optimization pass is typed to operate at exactly one of these granularities, and its placement within a pipeline must respect this hierarchy. For instance, a \texttt{LoopPass} can only appear within a \texttt{loop} manager, which itself must be contained inside a \texttt{function} manager.

This hierarchical model is realized through a system of pass managers and adaptors. Consequently, an optimization pipeline is no longer a simple, unstructured sequence. Instead, it assumes a specific, organized form based on how passes are grouped and nested, a configuration we refer to as a pipeline \textit{skeleton}. The skeleton determines the precise scope and context for each pass. This design provides several significant advantages over the legacy system. First, it enables a more refined and efficient \textit{analysis management} mechanism: analyses are computed lazily, cached, and invalidated with fine-grained precision, thereby avoiding unnecessary recomputation. For example, a pass that modifies a single function invalidates only the analyses pertaining to that function, leaving all others unaffected. Second, this structured grouping improves \textit{cache locality} during compilation. By executing multiple \texttt{FunctionPass}es consecutively on one function before moving to the next, the compiler can make better use of CPU caches. These characteristics show that the grouping and nesting of passes—the skeleton—is not only a syntactic requirement, but also a structural feature with direct implications for compilation efficiency and, as we will argue, for the quality of the generated code.

\subsection{Motivation}
\label{ssec:motivation}

The hierarchical design of the LLVM New Pass Manager introduces a fundamental challenge for auto-tuning that poses fundamental challenges to traditional sequence-based approaches, which cannot guarantee syntactic validity. The primary issue is one of \textbf{syntactic validity}: any generated optimization pipeline must conform to the strict nesting rules of the New PM's grammar. An auto-tuner that generates syntactically incorrect pass strings would simply fail at compile-time, rendering the entire search process ineffective. Approaches that treat the pipeline as a flat list cannot guarantee this conformance, making them fundamentally unsuitable. This establishes the baseline requirement: a modern auto-tuner must be built upon a formal representation capable of modeling and manipulating these hierarchical structures.

Beyond this primary challenge of validity, however, lies a more subtle question: does the specific choice of a valid hierarchical structure impact the quality of the final optimized code? To investigate this, we conducted an experiment to isolate the effect of the pipeline's structure while keeping the sequence of optimization passes fixed. We applied a fixed sequence of passes—\texttt{globalopt} (M), \texttt{inline} (C), \texttt{gvn} (F), and \texttt{loop-deletion} (L)—to several CBench programs using five distinct, but equally valid, nesting skeletons that preserve the pass order. These skeletons are defined as follows:

\begin{itemize}
\item \textbf{Fully Sequential:} \
\texttt{module(M), module(cgscc(C)), module(function(F)), \\ module(function(loop(L)))}
\item \textbf{F+L Combined:} \
\texttt{module(M), module(cgscc(C)), module(function(F, loop(L)))}
\item \textbf{M+C, F+L Combined:} \
\texttt{module(M, cgscc(C)), module(function(F, loop(L)))}
\item \textbf{C+F+L Combined:} \
\texttt{module(M), module(cgscc(C, function(F, loop(L))))}
\item \textbf{Fully Nested:} \
\texttt{module(M, cgscc(C, function(F, loop(L))))}
\end{itemize}

\begin{table*}[htbp]
\centering
\footnotesize
\caption{Instruction count variation due to skeleton structure for representative CBench programs where the impact is most notable. All tests use the fixed pass sequence (\texttt{globalopt, inline, gvn, loop-deletion}), with the lowest instruction count for each program highlighted in bold.}
\label{tab:skeleton_impact}
\begin{tabular}{l|c|ccccc}
\hline
\textbf{Program} & \textbf{Original IC} & \textbf{Skeleton 1} & \textbf{Skeleton 2} & \textbf{Skeleton 3} & \textbf{Skeleton 4} & \textbf{Skeleton 5} \\
 & & (Sequential) & (F+L Comb.) & (M+C, F+L Comb.) & (C+F+L Comb.) & (Fully Nested) \\
\hline
dijkstra    & 450  & \textbf{372} & \textbf{372} & \textbf{372} & 481 & 481 \\
patricia    & 1282 & \textbf{958} & \textbf{958} & \textbf{958} & 1042 & 1042 \\
qsort       & 638  & 627 & 627 & 627 & \textbf{601} & \textbf{601} \\
sha         & 799  & \textbf{687} & \textbf{687} & \textbf{687} & 758 & 758 \\
stringsearch & 1348 & 1109 & 1109 & 1109 & \textbf{1108} & \textbf{1108} \\
\hline
\end{tabular}
\end{table*}

As summarized in Table~\ref{tab:skeleton_impact}, the results demonstrate that for the same sequence of passes, different nesting structures can indeed lead to different performance outcomes. This behavior arises because the nesting structure dictates the precise execution order and scope of optimizations across different IR units (such as functions). For example, a flatter, sequential structure might apply one pass to all functions before starting the next, while a deeply nested structure applies a sequence of passes to each function individually before moving to the next. This can alter which intermediate program states are visible to subsequent passes, thereby enabling or disabling different optimization opportunities.

While the primary challenge for any New PM auto-tuner is generating valid pipelines, these findings reveal an important secondary dimension to the problem: the choice among valid structures is not trivial and can influence optimization quality. This observation solidifies the need for a framework that does more than simply generate \textit{one} valid pipeline. A truly effective approach must treat the hierarchical structure as a first-class, manipulable entity, enabling the exploration of the space of valid hierarchical configurations.

\section{Formal Representation of Pass Pipelines}
\label{sec:method_representation}

To address the validity challenge outlined in our motivation and to systematically explore the structured search space of the LLVM New Pass Manager, we first define a formal representation for valid pass pipelines. The conventional one-dimensional list does not fully capture the hierarchical and nested nature of the New PM, nor can it guarantee syntactic correctness. Therefore, we represent a pass pipeline using two complementary formalisms: a Context-Free Grammar that specifies its syntactic validity, and a logical \textit{forest} data structure that facilitates its algorithmic manipulation.

\subsection{A Formal Grammar for Pipeline Skeletons}
\label{ssec:representation_grammar}

We define the set of all syntactically valid pass pipelines using a Context-Free Grammar, $G = (V, \Sigma, R, S)$. This grammar captures the hierarchical and recursive structure of the LLVM New Pass Manager, ensuring that only well-formed pipelines are considered during tuning.

\begin{itemize}
    \item \textbf{Non-terminals $V$}:  
    $\{ \langle \text{Pipeline} \rangle, \langle \text{ModuleManager} \rangle, \langle \text{ModuleElement} \rangle, \langle \text{CGSCCManager} \rangle, \\
    \langle \text{CGSCCElement} \rangle, \langle \text{FunctionManager} \rangle, \langle \text{FunctionElement} \rangle, \langle \text{LoopManager} \rangle, 
    \\ \langle \text{LoopElement} \rangle,
    \langle \text{ModulePass} \rangle, \langle \text{CGSCCPass} \rangle, \langle \text{FunctionPass} \rangle, \langle \text{LoopPass} \rangle \}$
    
    \item \textbf{Terminals $\Sigma$}:  
    $\{ \texttt{"module("}, \texttt{"cgscc("}, \texttt{"function("}, \texttt{"loop("}, \texttt{")"}, \texttt{","}, \text{pass\_names} \}$
    
    \item \textbf{Start Symbol $S$}: $\langle \text{Pipeline} \rangle$
\end{itemize}

The production rules $R$ are defined as follows. The Kleene star ($*$) denotes zero or more comma-separated repetitions of an element.

\begin{enumerate}
    \item[\textit{R1.}] $\langle \text{Pipeline} \rangle \rightarrow \langle \text{ModuleManager} \rangle \mid \langle \text{ModuleManager} \rangle \texttt{,} \langle \text{Pipeline} \rangle$
    
    \item[\textit{R2.}] $\langle \text{ModuleManager} \rangle \rightarrow \texttt{module(} (\langle \text{ModuleElement} \rangle \texttt{,})^* \langle \text{ModuleElement} \rangle \texttt{)}$
    
    \item[\textit{R3.}] $\langle \text{ModuleElement} \rangle \rightarrow \langle \text{ModulePass} \rangle \mid \langle \text{ModuleManager} \rangle \mid  \\ \langle \text{CGSCCManager} \rangle \mid  \langle \text{FunctionManager} \rangle$
    
    \item[\textit{R4.}] $\langle \text{CGSCCManager} \rangle \rightarrow \texttt{cgscc(} (\langle \text{CGSCCElement} \rangle \texttt{,})^* \langle \text{CGSCCElement} \rangle \texttt{)}$
    
    \item[\textit{R5.}] $\langle \text{CGSCCElement} \rangle \rightarrow \langle \text{CGSCCPass} \rangle \mid \langle \text{CGSCCManager} \rangle \mid \langle \text{FunctionManager} \rangle$
    
    \item[\textit{R6.}] $\langle \text{FunctionManager} \rangle \rightarrow \texttt{function(} (\langle \text{FunctionElement} \rangle \texttt{,})^* \langle \text{FunctionElement} \rangle \texttt{)}$
    
    \item[\textit{R7.}] $\langle \text{FunctionElement} \rangle \rightarrow \langle \text{FunctionPass} \rangle \mid \langle \text{FunctionManager} \rangle \mid \langle \text{LoopManager} \rangle$
    
    \item[\textit{R8.}] $\langle \text{LoopManager} \rangle \rightarrow \texttt{loop(} (\langle \text{LoopElement} \rangle \texttt{,})^* \langle \text{LoopElement} \rangle \texttt{)}$
    
    \item[\textit{R9.}] $\langle \text{LoopElement} \rangle \rightarrow \langle \text{LoopPass} \rangle \mid \langle \text{LoopManager} \rangle$
\end{enumerate}

This grammar enforces the hierarchical constraints of the LLVM New Pass Manager. For example, a \texttt{module} manager (R3) can directly host \texttt{cgscc} and \texttt{function} managers, but \texttt{loop} managers are introduced only under \texttt{function} managers (R7). The recursive definitions (e.g., R5, R7, R9) allow managers to be nested at the same level, supporting staged optimizations over the same IR unit. Under these rules, pipelines generated from this grammar are syntactically valid.

\subsection{The Skeleton as a Forest Data Structure}
\label{ssec:representation_forest}

\begin{figure}[htbp]
\centering
\footnotesize
\begin{forest}
  for tree={
    draw, semithick, rounded corners,
    font=\sffamily\small,
    minimum width=2.5cm,
    node options={align=center},
    edge={-latex, semithick},
    s sep=10pt,
    l sep=5pt,
    tier/.style={for tree={l sep=12pt, s sep=3pt}},
    leaf/.style={fill=gray!20, rounded corners=2pt}
  }
[Pipeline (Forest)
  [ModuleManager (Tree)
    [ModulePass, leaf]
    [CGSCCManager
        [CGSCCPass, leaf]
        [FunctionManager
            [FunctionPass, leaf]
            [LoopManager, tier=loops
                [LoopPass, leaf]
            ]
        ]
    ]
  ]
  [ModuleManager (Tree)
    [FunctionManager
        [FunctionPass, leaf]
    ]
  ]
]
\end{forest}
\caption{Logical \textit{Forest} representation of a Pass Pipeline. Each root $\langle \text{ModuleManager} \rangle$ forms a tree. Non-leaf nodes are managers, and leaf nodes are terminal passes.}
\label{fig:grammar_forest}
\end{figure}
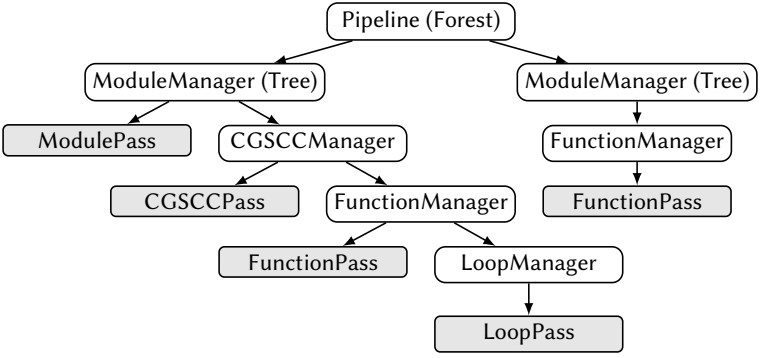

While the grammar defines the string representation, a logical data structure is required for algorithmic manipulation. We model a valid pipeline skeleton as a \textit{forest}—an ordered collection of trees, as illustrated in Figure~\ref{fig:grammar_forest}. Each tree represents an independent optimization stage, with manager nodes as internal nodes and terminal optimization passes as leaf nodes.

The structural components are defined as follows:

\begin{itemize}
    \item \textbf{Forest:} Corresponds to the grammar's start symbol $\langle \text{Pipeline} \rangle$ and consists of a list of \textit{trees}, representing sequential optimization phases.
    \item \textbf{Tree:} Rooted by a $\langle \text{ModuleManager} \rangle$, representing a single optimization stage.
    \item \textbf{Depth:} Maximum nesting level of managers within a tree, indicating the granularity of the optimization strategy.
    \item \textbf{Width:} Number of child nodes within a manager, representing the length and complexity of the optimization sequence at a given level.
\end{itemize}

These structural properties—forest size, tree composition, depth, and width—define the organization of a pipeline. For instance, a single, deep, narrow tree corresponds to a specialized optimization strategy, while a forest of shallow, wide trees corresponds to a phased, broader strategy. This structured representation serves as the foundation for knowledge-guided and evolutionary algorithms, enabling systematic manipulation of both pass selection and hierarchical arrangement.

\section{Methodology}
\label{sec:methodology}

\subsection{Framework Overview}
\label{sec:method_overview}

Navigating the hierarchical and syntactically constrained search space of the LLVM New Pass Manager presents challenges that motivate the need for a new auto-tuning approach. To address this, we propose a hybrid framework that combines offline knowledge mining with online, structure-aware evolutionary search. The design principle is to maintain syntactic validity at all stages while exploring the large space of pass configurations efficiently.

\begin{figure}[htbp]
  \centering
  \includegraphics[width=0.9\textwidth]{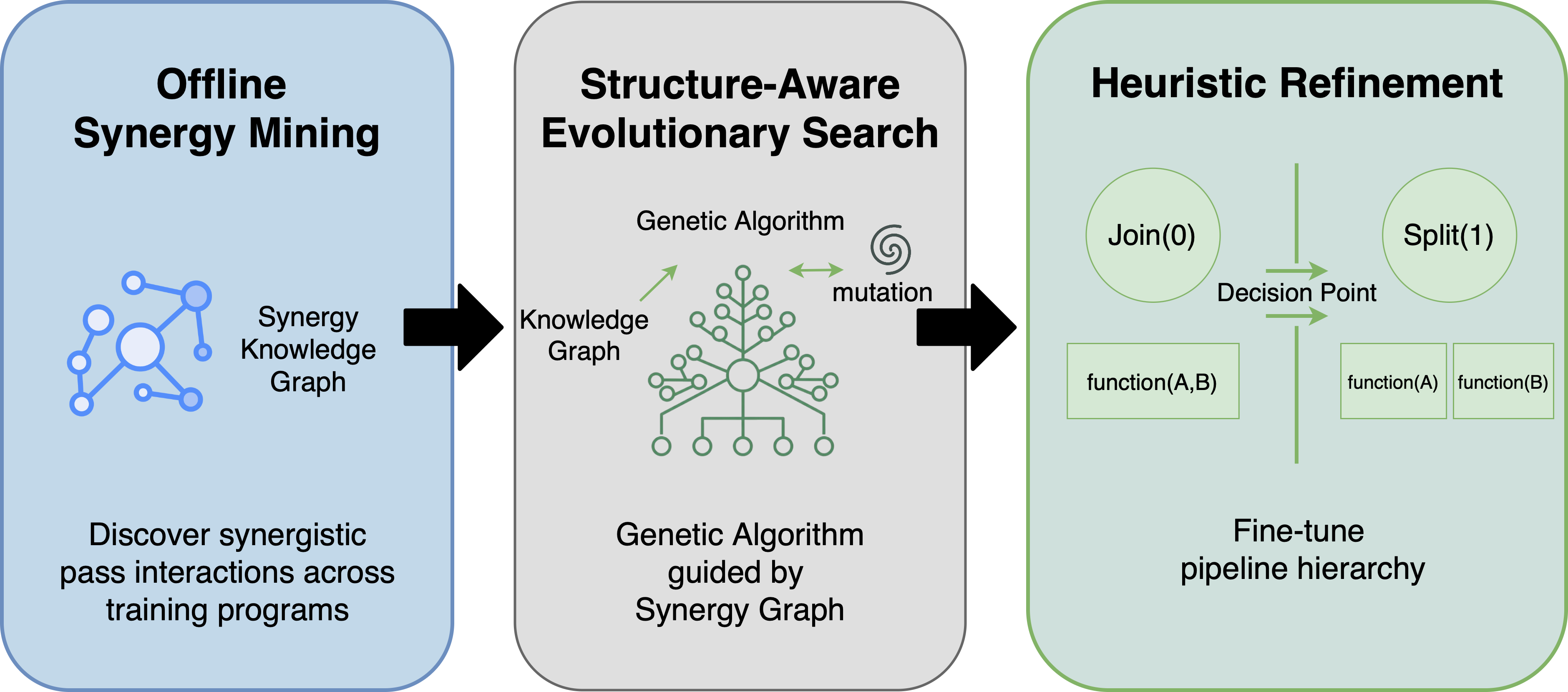}
  \caption{Overview of the auto-tuning framework. The offline mining produces a Synergy Knowledge Graph, which guides the generation of candidate pass sequences. Heuristic refinement is then applied to improve the hierarchical arrangement.}
  \label{fig:framework_overview}
\end{figure}

As illustrated in Figure~\ref{fig:framework_overview}, our framework consists of three stages that together form an integrated search strategy:
(1) \textbf{Offline Synergy Mining:} In a one-time offline phase, we construct a \textbf{Synergy Knowledge Graph} by analyzing pass interactions across a large corpus of programs. This graph encodes structure-aware relationships between passes, providing heuristic knowledge that improves the tractability of the subsequent online search.
(2) \textbf{Online Guided Evolutionary Search:} The core of the framework is a \textbf{structure-aware Genetic Algorithm (GA)}. Operating directly on the forest data structure defined in Section~\ref{sec:method_representation}, its genetic operators guarantee that every generated pipeline is syntactically valid. The search is guided by the Synergy Knowledge Graph to converge towards regions of the search space associated with better performance.
(3) \textbf{Optional Heuristic Refinement:} Finally, the framework includes a refinement stage. This step builds on the earlier observation that different valid nesting structures can affect performance, and makes localized heuristic adjustments to pipeline hierarchy to obtain modest additional improvements.

This multi-stage methodology decomposes the auto-tuning problem for the New PM into manageable components, aiming to balance syntactic validity with the potential for performance improvement.

\subsection{Offline Synergy Mining}
\label{sec:method_stage1_synergy_mining}

\subsubsection{Rationale for Synergy Mining}
\label{ssec:synergy_rationale}

The search space of valid pass pipelines in the LLVM New Pass Manager is combinatorially large, making a blind or purely random search for high-performance solutions infeasible. To address this, our framework employs a knowledge-guided approach that includes an offline phase to pre-mine beneficial pass interactions, or \textbf{synergistic pairs}. These serve two main purposes:
(1) \textbf{Guiding the Search Process:} A pre-computed knowledge base of synergistic pairs provides a heuristic for the online search. Rather than exploring the space uniformly, the search algorithm can prioritize sequences composed of passes that have been observed to interact positively. This focuses the search on regions of the space with higher expected performance, potentially improving search efficiency and the likelihood of discovering higher-performance pipelines.
(2) \textbf{Constructing Candidate Solutions:} Synergistic pairs act as building blocks informed by prior observations. This knowledge can be used to generate initial candidate solutions that incorporate known beneficial pass interactions. It can also inform modification operators (e.g., mutation), favoring changes that are more likely to lead to performance improvements, thereby enhancing the overall quality of solutions explored by the search.

As shown in Figure~\ref{fig:synergy_types}, because the New PM organizes passes hierarchically, we distinguish between two types of synergy: \textbf{Intra-Level Synergy}, between passes at the same level, and \textbf{Inter-Level Synergy}, between passes at different levels. Encoding this distinction in the knowledge graph allows the subsequent search to be both guided by prior observations and structure-aware.

\begin{figure}[htbp]
\centering
\begin{minipage}{0.5\textwidth}
    \centering
    \begin{forest}
      for tree={
        draw, semithick, rounded corners,
        font=\sffamily\small,
        node options={align=center},
        edge={-latex, semithick},
        s sep=6pt, l sep=12pt,
        where level=2{fill=blue!10, draw=blue!50, thick}{},
      }
    [ModuleManager
      [FunctionManager
        [instcombine, name=p1]
        [gvn, name=p2, xshift=70pt]
        [adce]
      ]
    ]
    \draw[<->, dashed, blue, thick] (p1.east) -- (p2.west) node[midway, above, font=\sffamily\tiny, black]{Sibling};
    \end{forest}
    \caption*{\centering(a) Intra-Level Synergy Example:\\ \texttt{module(function(instcombine, gvn))}}
\end{minipage}\hfill
\begin{minipage}{0.5\textwidth}
    \centering
    \begin{forest}
      for tree={
        draw, semithick, rounded corners,
        font=\sffamily\small,
        node options={align=center},
        edge={-latex, semithick},
        s sep=5pt, l sep=12pt,
      }
    [ModuleManager
      [globalopt, name=p1, fill=blue!10, draw=blue!50, thick]
      [FunctionManager
        [gvn,  xshift=-50pt, yshift=-20pt, name=p2, before computing xy={l=24pt}, fill=blue!10, draw=blue!50, thick]
        [adce, name=p3, yshift=-5pt, fill=blue!10, draw=blue!50, thick]
      ]
    ]
    \draw[->, dashed, blue, thick] (p1) -- (p2) node[midway, right, font=\sffamily\tiny, black]{Parent-Child};
    \draw[->, dashed, blue, thick] (p1) -- (p3) node[midway, right, font=\sffamily\tiny, black]{Parent-Child};
    \end{forest}
    \caption*{\centering(b) Inter-Level Synergy Example:\\ \texttt{module(globalopt, function(licm, adce))}}
\end{minipage}
\caption{Illustration of the two primary structural synergy types. Blue nodes highlight the pass pair under evaluation. (a) shows two \texttt{FunctionPass}es as siblings. (b) shows a \texttt{ModulePass} providing context for a \texttt{FunctionPass}.}
\label{fig:synergy_types}
\end{figure}

\subsubsection{Decoupling Sequence from Micro-Structure for Tractable Mining}
\label{ssec:synergy_decoupling}

A key challenge in analyzing pass interactions within the New PM is the entanglement of pass sequence and nesting structure. In this initial stage, the goal is to identify first-order synergistic relationships between passes that are empirically observed to affect performance. To this end, we temporarily decouple the pass sequence from its micro-level structural arrangement.

To validate this simplification, we systematically evaluated the impact of micro-structures. Even for a minimal two-pass sequence, the New PM allows multiple valid structural arrangements depending on the pass types. For our experiment, we considered the following representative cases:

Intra-Level pairs of two Function passes ($F_1, F_2$), three structural variants:
\begin{enumerate}
    \item \textbf{Micro:} tightly coupled within a single manager. Skeleton: \texttt{module(function($F_1, F_2$))}
    \item \textbf{Meso:} sequential within the same tree, sharing a module-level context. \\ Skeleton: \texttt{module(function($F_1$), function($F_2$))}
    \item \textbf{Macro:} completely independent optimization stages. Skeleton: \texttt{module(function($F_1$)), module(function($F_2$))}
\end{enumerate}

Inter-Level pairs of a Module pass ($M_1$) and a Function pass ($F_1$), two variants:
\begin{enumerate}
    \item \textbf{Nested:} both inside a single tree. Skeleton: \texttt{module($M_1$, function($F_1$))}
    \item \textbf{Phased:} placed in separate stages. Skeleton: \texttt{module($M_1$), module(function($F_1$))}
\end{enumerate}

To quantify the impact of micro-structures on pairwise synergies, we conducted a large-scale experiment on a subset of the training programs (used only for discovering synergistic pairs, not for final testing). We randomly sampled a large number of two-pass pairs, spanning both intra- and inter-level combinations, and evaluated each pair across all applicable structural variants. The results were consistent: in over 99.7\% of cases, different micro-structures for a given pass pair produced identical final instruction counts.

These results provide empirical evidence that for simple two-pass interactions, the dominant performance factors are the choice and order of the passes themselves, while the micro-level nesting structure appears to have minimal impact. A small fraction of cases (<0.3\%) showed minor variations. The observed trend supports a pragmatic divide-and-conquer strategy for our framework. Accordingly, the offline mining phase focuses on capturing primary, sequence-oriented synergistic effects, temporarily decoupled from the less impactful structural variable. More subtle and relatively infrequent performance effects of nesting structure are addressed in a later, dedicated refinement stage (Section~\ref{ssec:refinement}). This approach ensures that the framework first establishes a foundation based on pass sequence and then performs fine-grained structural adjustments where they may provide additional benefit.

\subsubsection{Approach: Systematic Discovery and Quantification}
\label{ssec:synergy_approach}

Our methodology for mining synergistic relationships is encapsulated in Algorithm~\ref{alg:synergy_discovery}. The design of this algorithm is directly informed by the findings in the preceding section. Based on the evidence that micro-structure appears to have limited impact on pairwise interactions, we adopt a unified and efficient approach. We systematically evaluate all considered two-pass combinations using a single, representative skeleton, regardless of whether the pair is intra-level or inter-level. This allows us to isolate and quantify the primary, sequence-oriented synergistic effects across the training programs.

\begin{algorithm}[htbp]
\caption{Synergy Knowledge Graph Construction}
\label{alg:synergy_discovery}
\begin{algorithmic}[1]
\Require Dataset $D$, PassDatabase $PDB$
\State $KB\_counts \gets \text{new defaultdict(list)}$
\ForAll{program $p \in D$}
    \State $IR \gets \text{load\_ir}(p)$; $IC_{orig} \gets \text{get\_instruction\_count}(IR)$
    \State $BaselinePerf \gets \text{calculate\_single\_pass\_performance}(IR, IC_{orig}, PDB)$
    \ForAll{structured pair $(P_1, P_2)$ from all passes in $PDB$}
        \State $S_{combined} \gets \text{build\_representative\_skeleton}(P_1, P_2)$ 
        \State $T \gets \text{get\_synergy\_type}(S_{combined})$
        \State $IC_{combined} \gets \text{evaluate}(IR, S_{combined})$
        \State $\text{Perf}_{combined} \gets IC_{orig} - IC_{combined}$
        \State $\text{Perf}_{sum} \gets BaselinePerf[P_1] + BaselinePerf[P_2]$
        \If{$\text{Perf}_{combined} > \text{Perf}_{sum}$}
            \State $KB\_counts[P_1.\text{name}].\text{append}(\{P_2.\text{name}, T\})$
        \EndIf
    \EndFor
\EndFor
\State $KnowledgeGraph, StartWeights \gets \text{normalize\_and\_build}(KB\_counts)$
\State \Return $KnowledgeGraph, StartWeights$
\end{algorithmic}
\end{algorithm}

\noindent
\textbf{Synergy Knowledge Graph:}  
\begin{itemize}
    \item \textbf{Node:} each node represents an individual compiler pass.  
    \item \textbf{Edge:} a directed edge $(P_1 \to P_2)$ indicates a positive synergy, i.e., $P_1$ followed by $P_2$ yields better performance than the sum of their individual effects.  
    \item \textbf{Edge type:} labeled as \textit{Intra-Level} (both passes in the same manager) or \textit{Inter-Level} (passes across different managers).  
    \item \textbf{Weight:} computed uniformly by counting occurrences of positive synergy across the dataset, then normalized across all outgoing edges from $P_1$.  
\end{itemize}

The algorithm first establishes a performance baseline for every individual pass. The main loop then iterates through all valid structured pairs. For each pair $(P_1, P_2)$, it constructs a single, representative, nested skeleton and evaluates its performance. The synergy is quantified by comparing the combined benefit against the sum of individual benefits. If a positive synergy is detected, the directed relationship from $P_1$ to $P_2$, including its structural type, is recorded . This recording is asymmetric to capture the order-dependent nature of pass interactions. After processing all programs, the aggregated counts are normalized to generate the final weighted Synergy Knowledge Graph and start-pass probabilities .

\subsection{Structure-Aware Evolutionary Search}
\label{sec:method_stage2_evolutionary_walk}

With the Synergy Knowledge Graph providing heuristic guidance, the next stage of our framework employs a search algorithm that can systematically explore the vast space of valid pipelines. The primary requirement for this algorithm is that it must operate natively on the hierarchical forest representation defined in Section~\ref{sec:method_representation} and guarantee that all operations preserve syntactic validity. To this end, we design a \textbf{structure-aware Genetic Algorithm (GA)} whose components are tailored to this constrained search space. Each individual in the GA is represented as a forest data structure corresponding to a complete and valid pipeline, and the search process is guided by the pre-computed knowledge graph $G_S = (V, E, W, T)$.

\subsubsection{Initial Population Generation}

The initial population is generated by performing a \textbf{Weighted Random Walk} on the knowledge graph $G_S$. This approach seeds the population with combinations of passes observed to be synergistic, providing an informed initialization for the subsequent search. For each individual, the process is as follows:

\begin{enumerate}
    \item \textbf{Start Pass Selection:} A starting pass $p_0$ is selected according to a probability distribution $D_{\text{start}}$. The probability $D_{\text{start}}(p_i)$ is proportional to the frequency with which pass $p_i$ was observed as an initiator of positive synergy during the offline mining phase.

   \item \textbf{Sequence Extension:} Given the last pass in the current sequence, $p_k$, the next pass $p_{k+1}$ is chosen from its set of successors $N(p_k)$ in $G_S$. The selection probability is proportional to the synergy weight:  
    $$ P(p_{k+1} | p_k) \propto W(p_k, p_{k+1}), \quad \forall p_{k+1} \in N(p_k) $$  
    
    The placement of $p_{k+1}$ within the forest structure is determined by the synergy’s edge type $T(p_k, p_{k+1})$. For example, an \textit{Intra-Level} synergy may place $p_{k+1}$ as a sibling to $p_k$ within the same pass manager, whereas an \textit{Inter-Level} synergy may introduce a new nested manager. This process continues until a predefined maximum sequence length is reached.  

\end{enumerate}

\subsubsection{Crossover and Mutation}
The crossover operator generates offspring by exchanging structural subtrees between two parent individuals. For each parent, a random subtree (defined as a manager and its contained passes) is selected. These two subtrees are swapped to produce two offspring. After the swap, the resulting individuals are validated against the syntactic rules; if a new structure is invalid, the crossover is discarded.

The mutation operator includes both addition and replacement operations. A random pass within the individual’s structure is selected as an \textit{anchor}. The algorithm queries the synergy knowledge graph to identify passes with known synergetic relationships to the anchor. One such partner is preferentially chosen for addition or replacement. If no suitable partner is found or the operation cannot be performed, the algorithm falls back to inserting or replacing with a randomly selected valid pass, thereby promoting structural diversity in the population.

\subsection{Heuristic Refinement}
\label{ssec:refinement}

In our methodology, we decoupled pass sequence from micro-structure during the initial knowledge mining phase (Section~\ref{ssec:synergy_decoupling}). We now return to this second optimization variable. As shown in Table~\ref{tab:skeleton_impact}, the hierarchical structure of a pipeline, even for a fixed pass sequence, can alter the final optimization result. This section examines the mechanism behind this effect and introduces the final stage of our framework, which is designed to systematically explore it.

\subsubsection{Impact of Nested Pipeline Structures}

Consider two function passes \texttt{A} and \texttt{B} (e.g., \texttt{InstCombine} and \texttt{GVN}) and three common ways of nesting them within a module pipeline:

\begin{enumerate}
    \item \texttt{module(function(A,B))}: Each function undergoes \texttt{A} followed by \texttt{B} sequentially.
    \item \texttt{module(function(A)), module(function(B))}: All functions in the module first execute \texttt{A}, then all execute \texttt{B}.
    \item \texttt{module(function(A), function(B))}: Semantically equivalent to the previous case; all functions first run \texttt{A}, then all run \texttt{B}.
\end{enumerate}

Assume a module with two functions, \texttt{foo} and \texttt{bar}, where \texttt{foo} calls \texttt{bar}. The execution order differs across nesting styles:

\begin{itemize}
    \item In \texttt{module(function(A,B))}, \texttt{foo} executes \texttt{A} and \texttt{B} before \texttt{bar} has been optimized. The transformation seen by \texttt{foo} depends on the unoptimized \texttt{bar}.
    \item In \texttt{module(function(A)), module(function(B))} or \texttt{module(function(A), function(B))}, all functions first execute \texttt{A}, so when executing \texttt{B}, \texttt{foo} sees the IR of \texttt{bar} after \texttt{A}. This can enable or disable certain optimizations compared to the first structure.
\end{itemize}

\noindent
\textbf{Observation:} The same set of passes, even when applied in the same order, can produce different optimization effects solely due to structural differences in the pipeline hierarchy. This indicates that pass execution is determined not only by the sequence but also by \emph{which IR states are visible to each pass at a given time}.

Figure~\ref{fig:nested-pass-execution} shows the two different execution styles in pseudocode form. The first code block corresponds to \texttt{module(function(A)), module(function(B))} or \texttt{module(function(A), function(B))}, and the second corresponds to \texttt{module(function(A,B))}.

\begin{figure}[htbp]
\centering

\begin{subfigure}{0.48\textwidth}
\begin{lstlisting}[language=Python, caption={All functions first run Pass A, then all functions run Pass B.}, label={subfig:pass-a-then-b}]
for f in module.functions:
    run Pass A on f
for f in module.functions:
    run Pass B on f
\end{lstlisting}
\end{subfigure}
\hfill 
\begin{subfigure}{0.48\textwidth}
\begin{lstlisting}[language=Python, caption={Each function runs Pass A immediately followed by Pass B.}, label={subfig:pass-a-and-b}]
for f in module.functions:
    run Pass A on f
    run Pass B on f
\end{lstlisting}
\end{subfigure}

\caption{Comparison of hierarchical pass execution strategies.}
\label{fig:nested-pass-execution}
\end{figure}

The analysis above shows how different nesting structures can create or miss optimization opportunities by altering the propagation of IR transformations. A search algorithm that only optimizes the pass sequence may settle on a solution with a sub-optimal structure. Therefore, the final stage of our framework is a dedicated \textbf{heuristic refinement process}. After the evolutionary search has identified a high-performance pass sequence, this stage performs targeted, localized adjustments to its nesting structure. The objective is to capture additional performance improvements by aligning the structure more effectively with inter-procedural dependencies.

\subsubsection{Methodology: Optimizing Pipeline Partitions via a Genetic Algorithm}
\label{ssec:refinement_methodology}

The analysis in Section~\ref{ssec:refinement} indicates that for a fixed pass sequence, the pipeline’s hierarchical structure is a critical optimization variable. We formalize this structural optimization as a sequence partitioning problem and employ a Genetic Algorithm (GA) to search for effective solutions. The methodology follows a four-stage process: modeling the problem, analyzing the search space, pruning it based on domain-specific knowledge, and finally, applying an algorithmic search.

\paragraph{Problem Modeling: The Partitioning Space}
Let $S = (p_1, p_2, \dots, p_N)$ be a fixed pass sequence of length $N$. A hierarchical structure for this sequence can be uniquely defined by a \textbf{partitioning} $\mathcal{P}$, which divides the sequence into a series of $m$ contiguous, non-overlapping blocks:
$$ \mathcal{P} = \{B_1, B_2, \dots, B_m\} $$
where $B_1 = (p_1, \dots), B_2 = (\dots), \dots, B_m = (\dots, p_N)$. Each block $B_j$ corresponds to a single pass manager in the final pipeline.

This partitioning is determined by the set of $N-1$ boundaries between adjacent passes. At each boundary $i \in \{1, \dots, N-1\}$, a binary decision is made: either to "split" the sequence (creating a new block) or to "join" it (extending the current block). The set of all possible partitions, $\mathbb{P}$, forms the search space. The size of this space is $|\mathbb{P}| = 2^{N-1}$, presenting a combinatorial optimization challenge. Our goal is to find an optimal partition $\mathcal{P}^*$ that maximizes a fitness function $f(\mathcal{P})$, which typically corresponds to the performance improvement over a baseline.
$$ \mathcal{P}^* = \arg\max_{\mathcal{P} \in \mathbb{P}} f(\mathcal{P}) $$

To intelligently navigate this large search space, we analyze the problem from its minimal unit: the structure governing two adjacent passes, $(p_i, p_{i+1})$. The performance implication of a "split" versus a "join" at boundary $i$ is the core of our analysis. This reveals that the nature of the boundary is the critical factor. We define $T(p)$ as the type of a pass (e.g., Function, Module). The key distinction is whether $T(p_i) = T(p_{i+1})$.

\paragraph{Rationale-Based Pruning of the Search Space}
This minimal unit analysis provides a powerful, domain-specific insight that allows for massive pruning of the search space. We identify two distinct cases for any boundary $i$:

\begin{enumerate}
    \item \textbf{Heterogeneous Boundary ($T(p_i) \neq T(p_{i+1})$):} Here, the two passes must reside in different manager types. While both "joined" (e.g.,\textit{ module(M, function(F))}) and "split" (e.g., \textit{module(M), module(function(F))}) forms are syntactically possible, their execution order is semantically equivalent due to LLVM's execution model. As this choice does not offer an optimization trade-off, these boundaries are not true variables. We can prune them from the search, treating them as constants where a "split" is always applied.

    \item \textbf{Homogeneous Boundary ($T(p_i) = T(p_{i+1})$):} Here, the "join" (e.g., \textit{function(A,B)}) and "split" (e.g., \textit{function(A), function(B)}) decisions lead to fundamentally different execution orders and, consequently, can directly impact performance. These boundaries are therefore the true variables in our optimization problem.
\end{enumerate}

This pruning strategy is the cornerstone of our refinement stage. It transforms the problem by identifying a small subset of boundaries, the \textbf{decision points}, which we denote by the set $D = \{i \mid T(p_i) = T(p_{i+1})\}$. The search is thus reduced from the full space $\mathbb{P}$ to a much smaller, more meaningful space of size $2^{|D|}$.

\paragraph{Algorithm: Genetic Search for Optimal Partitioning}
We formalize the pruned problem and apply a Genetic Algorithm (GA) to explore the space of candidate solutions. A partitioning strategy is encoded as a binary chromosome of length $k=|D|$, where each bit corresponds to a decision point and represents either a "join" (\textit{0}) or "split" (\textit{1}).

The GA evolves a population of these chromosomes. The process begins with an initial population that includes the structure from the main evolutionary stage alongside other random variants. The fitness of each chromosome is determined by decoding it into a full pass pipeline string and measuring its performance on the target program. Using standard operators—selection, crossover, and mutation applied to the decision points—the algorithm iteratively updates the population. This approach allows systematic exploration of the reduced search space and identification of high-performing structural configurations.

\section{Experiments}
\label{sec:experiments}

We evaluate our framework by addressing four Research Questions (RQs):

(RQ1) How does our framework's overall performance compare to LLVM -Oz?

(RQ2) How does it compare against other state-of-the-art auto-tuners?

(RQ3) Is the synergy-guided search beneficial?

(RQ4) What is the contribution of the final structural refinement stage?

\subsection{Experimental Setup}
\label{ssec:setup}

\paragraph{Datasets}
We use programs from the CompilerGym benchmark suite~\cite{CompilerGym}, partitioned into a training set of 19,603 programs for offline synergy mining and a disjoint test set of 335 programs for evaluation. The detailed composition is shown in Table~\ref{tab:dataset_composition}.

\begin{table}[htbp]
\centering
\caption{Dataset Composition detailing the number of programs for training and testing from each source dataset.}
\label{tab:dataset_composition}
\begin{tabular}{l l r r}
\toprule
\textbf{Type} & \textbf{Dataset} & \textbf{Train} & \textbf{Test} \\
\midrule
\multirow{6}{*}{Uncurated} & blas & 133 & 29 \\
& github-v0 & 7,000 & 0 \\
& linux-v0 & 4,906 & 0 \\
& opencv-v0 \cite{opencv} & 149 & 32 \\
& poj104-v1 \cite{poj} & 7,000 & 0 \\
& tensorflow-v0 \cite{tensorflow} & 415 & 90 \\
\midrule
\multirow{4}{*}{Curated} & cbench-v1 \cite{cbench} & 0 & 11 \\
& mibench-v1 \cite{mibench} & 0 & 40 \\
& chstone-v0 \cite{chstone} & 0 & 12 \\
& npb-v0 \cite{npb} & 0 & 121 \\
\midrule
\textbf{Total} & -- & \textbf{19,603} & \textbf{335} \\
\bottomrule
\end{tabular}
\end{table}

\paragraph{Baselines for Comparison}
We compare our framework against two categories of baselines:
\begin{itemize}
    \item \textbf{LLVM -Oz:} A standard, highly-engineered optimization level.
    \item \textbf{Auto-Tuning Methods:} We evaluate several representative auto-tuning methods originally designed for linear pass sequences: OpenTuner~\cite{opentuner}, GA~\cite{GA}, CFSAT~\cite{cfsat}, TPE~\cite{TPE}, CompTuner~\cite{Comptuner}, RIO~\cite{RIO}, and BOCA~\cite{BOCA}.
\end{itemize}

\paragraph{Evaluation Metric}
Our primary metric is the \textbf{Average OverOz (\%) Improvement}, which measures the additional percentage reduction in LLVM IR instruction count relative to the \texttt{opt -Oz} baseline. It is calculated as:
$$ \text{OverOz (\%)} = \frac{1}{|\mathcal{P}_\text{test}|} \sum_{p \in \mathcal{P}_\text{test}} \frac{I_{\text{Oz}}(p) - I_{\pi^*}(p)}{I_{\text{Oz}}(p)} \times 100 $$
where $\mathcal{P}_\text{test}$ denotes the set of test programs (with $|\mathcal{P}_\text{test}|$ being the number of test programs), $I_{\text{Oz}}(p)$ is the instruction count after \texttt{-Oz}, and $I_{\pi^*}(p)$ is the count from the evaluated method $\pi^*$. A higher value indicates a more effective optimization.

\paragraph{Environment}
Our framework is built using LLVM 18.1.6. We use the \texttt{opt} tool to apply optimization pipelines and extract instruction counts. All experiments were conducted on a server with an AMD EPYC 7763 64-core processor. To ensure comparability across methods, all baseline auto-tuning methods operate on the same search space consisting of the  transformation passes extracted from this LLVM version. This uniform search space ensures that all methods are evaluated under the same conditions.

\subsection{RQ1 \& RQ2: Overall Performance Comparison}

To address our first two research questions, we conduct a comprehensive performance evaluation of our framework against both the industrial-strength LLVM \texttt{-Oz} baseline (RQ1) and a suite of state-of-the-art auto-tuning methods (RQ2).

\paragraph{Results}
Table~\ref{tab:rq1_results} presents the performance of all evaluated methods, measured by the Average OverOz (\%) Improvement metric. \textbf{Answering RQ1}, our framework shows a notable and consistent performance improvement over the highly-tuned \texttt{opt -Oz} baseline, achieving an average additional instruction count reduction of \textbf{13.62\%}. \textbf{Answering RQ2}, the comparison against other auto-tuners reveals a stratification of performance levels. Our framework achieves higher performance than all other methods across the datasets. CFSAT follows as the second-best method with an average improvement of 8.27\%. OpenTuner achieves a slightly positive average (0.28\%) but with inconsistent results. In contrast, the remaining baselines (GA, TPE, and RIO) show average performance degradation, suggesting that they may be less suitable for this complex tuning task. Moreover, our framework achieves an average tuning time of only \textbf{5s}, slightly faster than CFSAT’s 6s.

\begin{table}[htbp]
\centering
\caption{Overall performance comparison based on Average OverOz (\%) Improvement. A higher value is better. The best result for each dataset is highlighted in bold. Results for some baselines are pending (--).}
\label{tab:rq1_results}
\resizebox{\textwidth}{!}{
\begin{tabular}{l c c c c c c c c}
\toprule
\textbf{Dataset} & \textbf{Ours} & \textbf{CFSAT} & \textbf{OpenTuner} & \textbf{GA} & \textbf{TPE} & \textbf{RIO} & \textbf{CompTuner} & \textbf{BOCA} \\
\midrule
cbench & \textbf{11.84} & 8.07 & -24.07 & -34.65 & -115.86 & -81.64 & -121.42 & -83.98 \\
blas & \textbf{6.86} & 3.92 & 0.47 & -18.19 & -20.86 & -19.04 & -21.35 & -18.20 \\
opencv & \textbf{8.02} & 7.42 & 7.82 & -8.93 & -13.83 & -13.47 & -13.85 & -13.28 \\
mibench & \textbf{7.99} & 5.56 & -0.71 & -15.88 & -146.54 & -40.64 & -161.58 & -158.20 \\
chstone & \textbf{21.03} & 12.67 & 2.48 & -42.27 & -102.80 & -68.95 & -108.94 & -87.47 \\
tensorflow & \textbf{8.05} & 7.64 & 6.90 & -9.57 & -14.36 & -14.60 & -15.13 & -14.01 \\
npb & \textbf{31.58} & 12.62 & 9.06 & 11.14 & -36.97 & -5.47 & -50.01 & -28.53 \\
\midrule
\textbf{Average} & \textbf{13.62} & 8.27 & 0.28 & -16.91 & -64.46 & -34.83 & -70.32 & -42.10 \\
\bottomrule
\end{tabular}
}
\end{table}

\paragraph{Analysis of Results}
The performance differences between the methods can be attributed to their fundamental design choices regarding the search space and optimization strategy, particularly in the context of the LLVM New PM.

The limited performance of most baseline auto-tuners (GA, TPE, etc.) can be explained by several factors. \textbf{(1)} Their search algorithms are designed primarily for linear sequences and do not natively respect the New PM’s hierarchical grammar. As a result, many of the pass sequences they generate are syntactically invalid and rejected by the compiler, reducing the efficiency of their search. \textbf{(2)} These methods tend to focus on pass \textit{selection} rather than precise \textit{ordering}, thereby missing optimization opportunities that depend on sequence interactions. \textbf{(3)} The search space of  passes includes many detrimental passes that can significantly increase instruction count. Without effective mechanisms to avoid these, GA and TPE often incorporate them, leading to performance degradation.

By contrast, the stronger performance of our framework and CFSAT reflects their shared use of knowledge-driven design. \textbf{(1)} Both approaches consider pass selection and ordering simultaneously. \textbf{(2)} Both employ an offline knowledge extraction phase, which helps identify and exclude harmful passes, substantially reducing the effective search space.

The performance advantage of our framework over CFSAT underscores the important role of native structural awareness. CFSAT's knowledge base implicitly contains structurally valid pass combinations, as any invalid combinations would have been rejected by the compiler during its offline mining. However, its search algorithm remains fundamentally linear and cannot explore beyond simple sequential compositions of these valid blocks. Our framework, being natively hierarchical, can explore the full, rich space of complex nested structures (e.g., \textit{module(passA, function(passB, passC))}). This unique ability to discover and exploit structure-dependent optimization opportunities, which are invisible to CFSAT's linear search, helps to explain the observed performance differences.

\subsection{RQ3 \& RQ4: Ablation Studies on Framework Components}

To understand the individual contributions of our key design choices, we conduct two distinct ablation studies. The first (RQ3) evaluates the impact of the Synergy Knowledge Graph, while the second (RQ4) quantifies the value added by the final structural refinement stage.

\subsubsection{RQ3: The Impact of Synergy-Guided Search}

\paragraph{Experimental Design}
To isolate the effect of our knowledge-guided approach, we create a baseline called \textbf{Random-GA}. This variant uses the exact same structure-aware Genetic Algorithm as our main framework but is knowledge-blind: its initial population and subsequent mutations are generated by randomly selecting passes from the full set of  available passes. To specifically test the impact on search efficiency, we run both our guided framework and Random-GA with a reduced computational budget (smaller population and fewer generations).

\paragraph{Results and Analysis}
Table~\ref{tab:rq3_ablation} shows that our synergy-guided approach tends to achieve better performance than the Random-GA across all datasets. While our method still reaches a positive average improvement over \texttt{-Oz} even with a reduced budget, the Random-GA results in notable performance degradation. Specifically, the average improvement across all datasets is 0.76\% for our synergy-guided approach versus -8.29\% for Random-GA. This contrast suggests that heuristic guidance plays an important role in navigating such a vast and complex search space. Although Random-GA is structurally-aware, the large number of available passes leads it to often select detrimental ones. In comparison, the synergy-guided approach directs the search more effectively: by seeding the population with higher-quality building blocks and guiding mutations towards beneficial changes, it converges more rapidly to higher-performing regions. These findings provide evidence that the Synergy Knowledge Graph is a valuable component of our framework.

\begin{table}[htbp]
\centering
\caption{Ablation study for RQ3, comparing Synergy-Guided GA with a Random GA under a reduced search budget. Performance is measured by Average OverOz (\%) Improvement.}
\label{tab:rq3_ablation}
\begin{tabular}{l rr}
\toprule
\textbf{Dataset} & \textbf{Ours (Synergy-Guided)} & \textbf{Random-GA (Unguided)} \\
\midrule
cbench & -0.93 & -13.67 \\
mibench & -8.75 & -18.82 \\
chstone & -1.36 & -15.65 \\
blas & -2.39 & -6.63 \\
opencv & 1.86 & -3.58 \\
tensorflow & -0.46 & -3.70 \\
npb & 8.46 & -3.96 \\
\midrule
\textbf{Average} & \textbf{0.76} & \textbf{-8.29} \\
\bottomrule
\end{tabular}
\end{table}

\subsubsection{RQ4: The Contribution of the Structural Refinement Stage}

\paragraph{Experimental Design}
This ablation study quantifies the value added by the final structural refinement stage. We compare the performance of our framework's main search stage (\textbf{Main GA}, before refinement) against the final output of the \textbf{Full Framework}.

\paragraph{Results and Analysis}
Table~\ref{tab:rq4_refinement} shows that the refinement stage provides a modest performance improvement, increasing the average OverOz score from 13.08\% to 13.62\%. Notably, the gain is totally pronounced on the \texttt{npb} dataset (+3.83\%) while no improvements at all on other dataset. This behavior also proves that the structures of the pass sequence have different infuluences on different programs.

NPB (NAS Parallel Benchmarks) primarily includes programs related to scientific and parallel computing. These programs typically contain a large number of loop structures (for/while) with deep nesting and high iteration counts. The code size of NPB programs therefore are highly dependent on loop optimizations. The nested structure may enable loop passes and function passes to collaborate effectively, fully leveraging the benefits of loop optimizations with the intra-procedural information obtained from the function-level passes. Thus, the choice of nesting structures can lead to substantial divergence in the instruction counts reduction. For other datasets, they contain fewer loop structures or shallower loop nesting, limiting the overall impact of loop optimizations on instruction counts. The code logic may be more dispersed. They are primarily I/O-, branch-, or system call-intensive, with bottlenecks not residing within loops, leaving little room for additional optimization from nested structures. Thus, the nested structure optimization may have a limited effect on the final instruction counts for these datasets.

These results provide evidence that our approach demonstrates good performance in the cases where programs are parallel and loop-intensive, with the refinement stage acting as a specialized component that can capture additional performance gains on benchmarks where pipeline structure becomes a significant secondary factor.

\begin{table}[htbp]
\centering
\caption{Ablation study for RQ4, quantifying the gain from the structural refinement stage. Performance is measured by Average OverOz (\%) Improvement.}
\label{tab:rq4_refinement}
\begin{tabular}{l rr r}
\toprule
\textbf{Dataset} & \textbf{Main GA} & \textbf{Full Framework} & \textbf{Gain from Refinement} \\
\midrule
cbench & 11.84 & 11.84 & +0.00\% \\
blas & 6.86 & 6.86 & +0.00\% \\
opencv & 8.02 & 8.02 & +0.00\% \\
mibench & 7.99 & 7.99 & +0.00\% \\
chstone & 21.03 & 21.03 & +0.00\% \\
tensorflow & 8.05 & 8.05 & +0.00\% \\
npb & 27.75 & \textbf{31.58} & \textbf{+3.83\%} \\
\midrule
\textbf{Average} & 13.08 & \textbf{13.62} & \textbf{+0.54\%} \\
\bottomrule
\end{tabular}
\end{table}

\subsection{Case Study: Structure as a Disambiguation Mechanism}

Our quantitative results have demonstrated the performance impact of structural choices. Here, we present a case study on \texttt{npb-v0\_6.ll} that illustrates an additional role of the hierarchical pipeline: its ability to resolve pass ambiguity, which can be problematic for flat-sequence representations. This case highlights a second mechanism through which structure influences optimization, complementing the execution-order effect discussed earlier.

\paragraph{The Limitation of Flat Sequences: The Ambiguity Problem}
A challenge for flat-sequence auto-tuners arises from passes with identical names but different functions depending on their scope. A representative example is \texttt{invalidate<all>}, a meta-pass that invalidates analyses at a specific IR level (Module, CGSCC, Function, or Loop). In a flat sequence, the string "\texttt{invalidate<all>}" does not specify scope, making its intended behavior unclear. As a result, a flat-sequence tuner may either omit such passes or adopt a fixed default interpretation, which might not be optimal.

\paragraph{Hierarchical Structure as the Solution}
By representing pipelines hierarchically, our framework removes this ambiguity: the nesting explicitly determines the pass’s scope. This can be seen in the two pipelines discovered for \texttt{npb-v0\_6.ll}. While they contain the same sequence of pass names, their different structures yield different semantics and thus different performance outcomes.
The initial pipeline from the main GA was a multi-stage, flattened sequence:
\begin{lstlisting}[language=bash, basicstyle=\footnotesize\ttfamily, breaklines=true, breakindent=0pt]
module(scc-oz-module-inliner),module(function(scalarize-masked-mem-intrin,jump-threading,correlated-propagation,reassociate,slp-vectorizer,vector-combine,correlated-propagation)),module(function(tsan,jump-threading,gvn-hoist,bounds-checking,instsimplify)),module(function(memcpyopt,adce)),module(function(loop(invalidate<all>))),module(strip)
\end{lstlisting}
Here, the nesting function(loop(...)) indicates that \texttt{invalidate<all>} is executed at the Loop level.
The refined pipeline, however, converged to a monolithic, nested structure:
\begin{lstlisting}[language=bash, basicstyle=\footnotesize\ttfamily, breaklines=true, breakindent=0pt]
module(scc-oz-module-inliner,function(scalarize-masked-mem-intrin,jump-threading,correlated-propagation,reassociate,slp-vectorizer,vector-combine,correlated-propagation,function(tsan,jump-threading,gvn-hoist,bounds-checking,instsimplify),memcpyopt,adce),invalidate<all>,strip)
\end{lstlisting}
In this case, since \texttt{invalidate<all>} is a direct child of the module manager, its scope is defined as the Module level.

\paragraph{Analysis: A Different Form of Structural Impact}
This example illustrates a type of structure effect distinct from the one shown in our motivation (Table \ref{tab:skeleton_impact}). In the earlier case, the effect came from altering the execution order of a fixed set of passes while their semantics remained unchanged. Here, however, the pass sequence is identical, yet the structural change redefines the semantics of \texttt{invalidate<all>}: one pipeline applies repeated loop-level invalidations, while the other performs a single module-level invalidation. These strategies impact the compiler’s analysis infrastructure in different ways, resulting in the observed 3.31\% performance gap. For this program, the refinement stage discovered that the module-level invalidation strategy was more effective.

\section{Limitations}  
Despite the effectiveness of our framework, there are several limitations.  
1) \textbf{Scope of Synergy Consideration:} Our current approach only accounts for intra-level and inter-level pass synergies, ignoring potential interactions between subtrees. Certain passes may form a high-quality combination that performs well across many programs. Treating such combinations as a subtree and exploring its synergy with other subtrees could further reduce the search space and potentially improve performance.  
2) \textbf{Refinement Adaptivity:} While the pass sequence itself dominates optimization performance in most cases, the nesting structure can still affect results for a small subset of programs. Our current refinement stage is not adaptive and cannot automatically identify which programs or pass sequences may benefit from structural adjustments.  
3) \textbf{Evaluation Metric:} We primarily use instruction count as the optimization objective. While widely adopted, it does not always reflect actual runtime performance or other metrics such as energy efficiency. Therefore, our pipelines may not be globally optimal under all practical performance criteria.  

Addressing these limitations, such as incorporating subtree-level synergies, developing adaptive refinement strategies, and considering more comprehensive evaluation metrics, represents promising directions for future work.

\section{Related Work}
\label{chap:related_work}

The automation of compiler optimization, particularly pass selection and phase ordering, has been a central theme of systems research for decades. Approaches have evolved through several distinct paradigms, each employing increasingly sophisticated techniques to navigate the vast optimization search space. We summarize these prior paradigms in Table~\ref{tab:related_work_summary}.

\begin{table}[h!]
\centering
\caption{Comparative Summary of Prior Auto-Tuning Paradigms.}
\label{tab:related_work_summary}
\begin{tabular}{l p{5cm} l}
\toprule
\textbf{Paradigm} & \textbf{Core Methodology} & \textbf{Representation} \\
\midrule
\textbf{Iterative Compilation} & Empirically searches the optimization space using heuristics (e.g., GA, Simulated Annealing) on a per-program basis. & Linear Sequence \\
\cite{bodin1998iterative, GA, cfsat, jantz2013exploiting, ICMC, BOCA, Comptuner, RIO, Reaction_Matching, opentuner} & & \\
\midrule
\textbf{ML / Reinforcement Learning} & Learns a predictive policy or cost model to map program features to effective optimizations, avoiding exhaustive search. & Linear Sequence \\
\cite{leather2020machine, autophase, CompilerGym, coreset, ashouri2018surveyauto-tuning, wang2018machine, deng2024compilerdream, shahzad2024neural} & & \\
\midrule
\textbf{Large Language Models} & Fine-tunes a foundation model to infer optimization strategies directly from source code, capturing deep semantic patterns. & Linear Sequence \\
\cite{cummins2024metalargelanguagemodel, pan2025compiler, cummins2023large, italiano2024finding} & & \\
\bottomrule
\end{tabular}
\end{table}

As summarized in the table, a critical thread across all prior paradigms—from heuristic search to machine learning and LLMs—is their foundational reliance on a \textbf{linear representation of the optimization pipeline}. This modeling choice, while perfectly suitable for legacy compiler infrastructures, is fundamentally incompatible with the hierarchical and grammatically-constrained architecture of the LLVM New Pass Manager. The core issue is that a linear representation lacks the expressive power to describe nested structures. Consequently, these methods are incapable of guaranteeing the generation of syntactically valid pipelines, which prevents them from being directly and reliably applied to the New PM's constrained search space. Furthermore, they are inherently "structurally-blind," meaning they cannot reason about or explore the performance implications of different valid hierarchical arrangements.

Our work addresses this \textbf{gap}. We are the first to propose an auto-tuning framework designed natively for the challenges and opportunities of modern, hierarchical compilers. By introducing a formal grammar and a forest data structure, we provide a robust mechanism for representing and manipulating these complex pipelines. Our knowledge-guided, co-evolutionary genetic algorithm treats the pipeline structure as a first-class entity. This ensures that every candidate solution is valid by construction and enables a holistic search that optimizes both the pass sequence and its hierarchical structure simultaneously, unlocking performance opportunities that are inaccessible to previous-generation, linear auto-tuners.

\section{Conclusion}
\label{sec:conclusion}
In this work, we presented a comprehensive framework for auto-tuning compiler passes in the LLVM New Pass Manager, addressing both the syntactic and structural challenges introduced by its hierarchical design. We first formalized the space of valid optimization pipelines using a context-free grammar and a forest-based data structure, providing a foundation for algorithmic manipulation that guarantees syntactic correctness. Building upon this representation, we introduced a structure-aware Genetic Algorithm guided by an offline-mined Synergy Knowledge Graph, enabling the efficient discovery of high-performance pass sequences. Furthermore, we designed an optional heuristic refinement stage to capture subtle performance differences caused by varying valid nesting structures. Our evaluation on seven benchmark datasets using LLVM 18.1.6 demonstrates that the proposed framework can achieve an average of 13.62\% additional reduction in instruction count compared to the standard \texttt{opt -Oz} optimization level, outperforming existing auto-tuning approaches. These results validate that integrating structural awareness, knowledge-guided search, and targeted refinement provides tangible benefits over both traditional linear-sequence auto-tuners and standard compiler optimization levels.

In summary, this work highlights the importance of considering hierarchical structure as a first-class factor in compiler auto-tuning and provides a general methodology for systematically exploring and exploiting this complex, constrained search space.

\section{Data Availability}
A replication package, including code and data used in our experiments, is available at \url{https://github.com/Mind4Compiler/NPMCT/} for review. Upon acceptance, we intend to make the data fully publicly available under an appropriate license, subject to anonymization and privacy constraints.

\bibliographystyle{ACM-Reference-Format}
\bibliography{ref}

\end{document}